\documentclass[doublecol]{epl2} 
\usepackage{geometry}
\usepackage{color}
\usepackage[normalem]{ulem}
\def\Fbox#1{\vskip1ex\hbox to 8.5cm{\hfil\fboxsep0.3cm\fbox{%
  \parbox{8.0cm}{#1}}\hfil}\vskip1ex\noindent}  %%  {TEXT} in BOX

\newcommand{\B}[1]{{\bm{#1}}}%% Bold Roman & Greek Lower & Upper Case
\let \= \equiv \let\*\cdot \let\~\widetilde \let\-\overline

\begin{document}

\title{Modeling Barkhausen Noise in Magnetic Glasses with Dipole-Dipole Interactions}
\author{Awadhesh K. Dubey\inst{1} \and H. George E. Hentschel\inst{2} \and Prabhat K. Jaiswal\inst{1} 
\and Chandana Mondal\inst{1} \and Itamar Procaccia\inst{1}\thanks{E-mail: \email{itamar.procaccia@gmail.com}} \and Bhaskar Sen Gupta\inst{1}
}
\shortauthor{Awadhesh K. Dubey \etal}
\institute{                    
  \inst{1} Department of Chemical Physics, Weizmann Institute of Science, Rehovot 76100, Israel\\
  \inst{2} Department of Physics, Emory University, Atlanta Ga.
}

\abstract{Long-ranged dipole-dipole interactions in magnetic glasses give rise to magnetic domains having labyrinthine patterns. Barkhausen Noise is
then expected to result from the movement of domain boundaries which is supposed to be modeled by the motion of elastic membranes with random pinning. We propose an atomistic model of such magnetic glasses in which
we measure the Barkhausen Noise which indeed results from the movement of domain boundaries. Nevertheless the statistics of
the Barkhausen Noise is found in striking disagreement with the expectations in the literature. In fact we find exponential statistics without any power law, stressing the fact that Barkhausen Noise can belong to very different universality classes. In this glassy system the essence of the phenomenon is the ability of spin-carrying particles to move and minimize the energy without any spin flip.  A theory is offered in
excellent agreement with the measured data without any free parameter.}

\maketitle
\begin{figure}[h!]
\centering
\includegraphics[scale = 0.45]{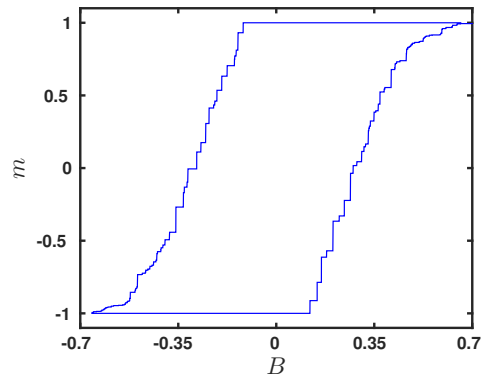}
\caption{A typical hysteresis loop showing the sharp changes in magnetization when a magnetic field in the $z$ direction is ramped
first up until saturation ($m=1$) and then down until saturation with $m=-1$. Our interest in this Letter is in the
statistics of the sharp changes $\Delta m$ seen in this figure.}
\label{hyst}
\end{figure}

%%%%%%%%%%%%%%%%%%%%%%%%%%%%%%%%%%
{\bf Introduction}: The statistics of so-called ``Serrated Noise" is a subject of wide-ranging interest from earthquakes with stress fluctuations on a global scale to Barkhausen Noise in small magnetic samples with magnetization jumps that are barely measurable. Typically one finds in such problems a wide range of sharp variations in some measurable quantity, and the question is how to model the statistics of these variations. In this Letter we return to Barkhausen Noise which is one of the most studied examples of serrated responses since its discovery in 1919 \cite{19Bar}. The phenomenon is manifested as a series of jumps in the magnetization of a ferromagnetic sample when subjected to varying external magnetic field \cite{69Bit, 76Bit,76MS,91CM,95PDS,96SBMS}. The phenomenon has practical importance for magnetic recordings \cite{92BZ} and for noninvasive material characterization \cite{94Sip}. When the magnetic field is ramped up and then down the magnetization describes a hysteresis loop
which is however punctuated by sharp jumps in measured value, see for example Fig. \ref{hyst}.
%%%%%%%%%%%%%%%%%%%%%%%%%%%%%%%

Some of the more careful experimental realizations of Barkhausen Noise involve magnetic systems with labyrinthine magnetic domains
in which the serrated response can be linked to the movement of the domain boundaries \cite{09DMW,08Col,06DZ}. In these cases theory
was proposed using a model of an elastic membrane that is pinned by random impurities and is moving under the action of a force. In their excellent review of both experiments and theory Durin and Zapperi \cite{06DZ} warn the reader that even in these well chosen
experiments the interpretation of the results is far from obvious, not the least because the statistics of Barkhausen Noise is not
invariant along the magnetization hysteresis loop. In less well characterized experiments Barkhausen Noise appears to be a very complex physical phenomenon with many different appearances. Its character may depend on the type of ferromagnetic specimen under
study, the character of the disorder in the material, the external field driving rate, thermal effects, strength of the demagnetization fields, and other experimental details.

\begin{figure}
\centering
\includegraphics[scale = 0.65]{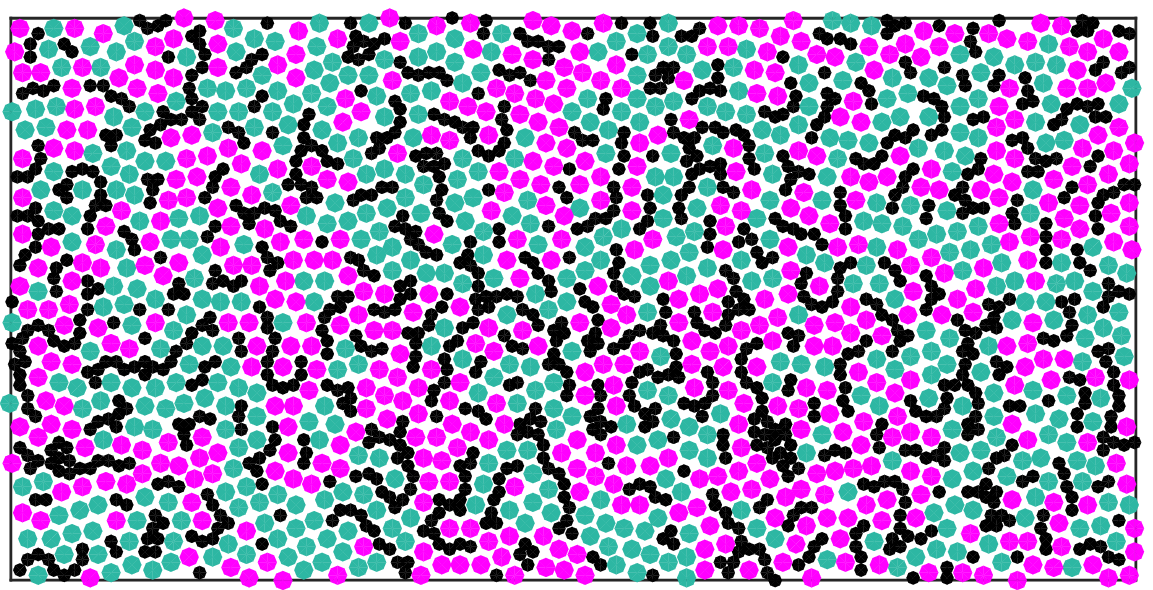}
\vskip 0.2 cm
\includegraphics[scale = 0.65]{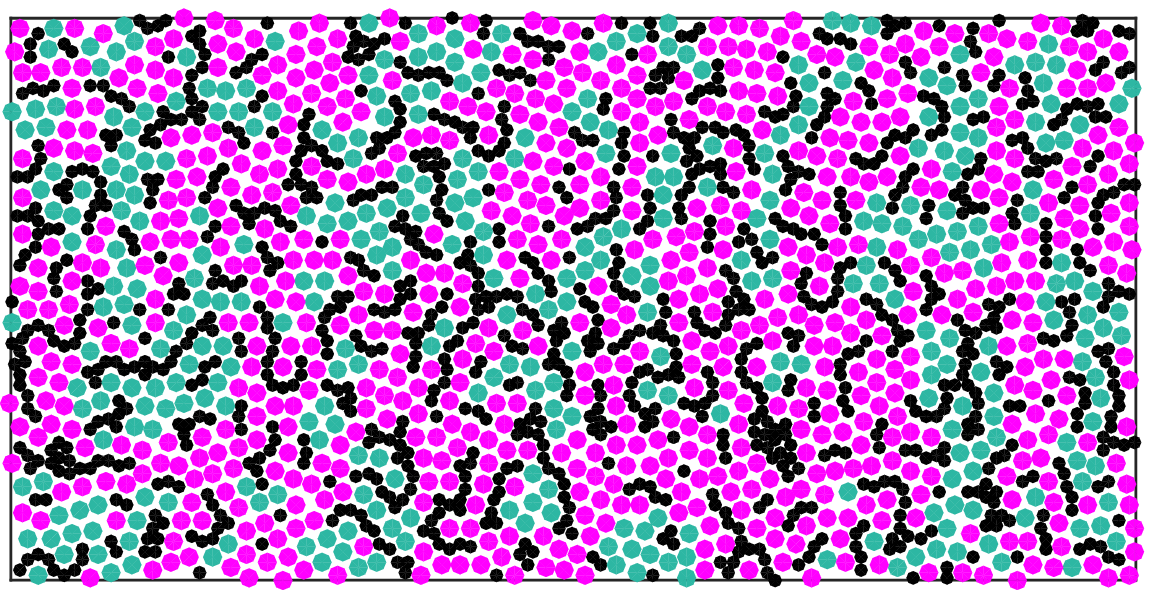}
\vskip 0.2 cm
\includegraphics[scale = 0.65]{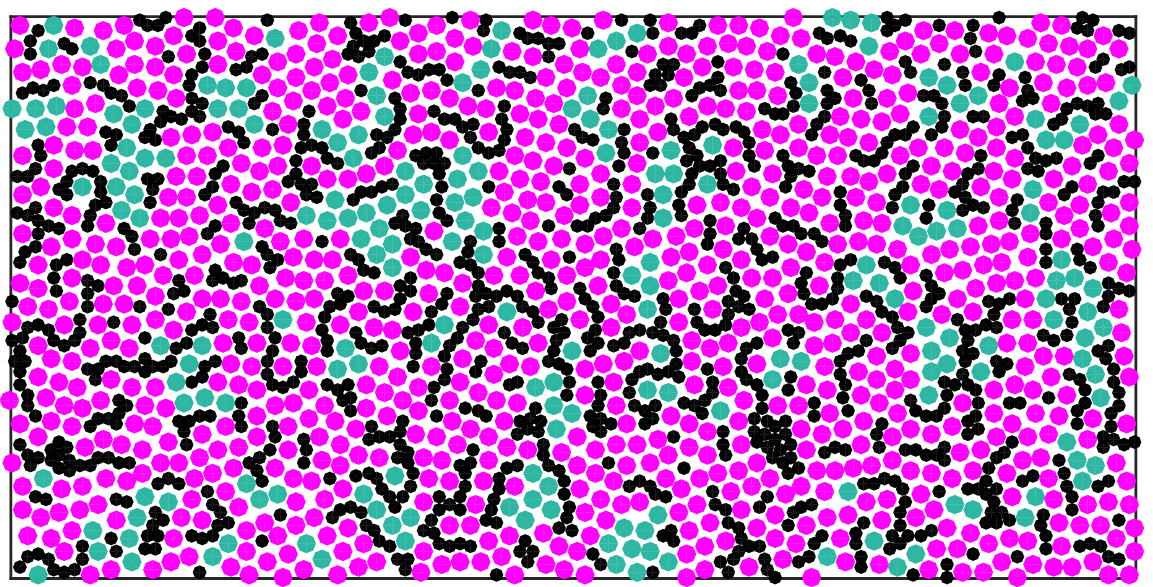}
\caption{A typical labyrinthine structure of the magnetic domains in the present model with $N=2100$ and dipole-dipole interactions. Here $J_0=6$, $K^\perp=0.25$, $K^\parallel =5$, $\mu^2_B=0.1$.
Upper panel: zero magnetic field. The ``small" particles in black are the non-magnetic particles.  The other two colors represent 50\% spins ``up" and 50\% spins ``down". Middle and lower panels: the effect of an increasing magnetic field on the labyrinthine pattern shown in the upper panel. Middle panel: $B=0.3$. Lower panel: $B=0.4$. Note the coarsening of the domains of ``up" spins (in magenta) which occur via movements of domain boundaries.}
\label{laby}
\end{figure}

For these reasons it is worthwhile to construct {\em microscopic} theoretical models in which the measurement can be done with arbitrary
accuracy and in which the interpretation can be fully justified by comparing careful simulations with the appropriate theory. Indeed,
in recent papers we initiated the microscopic study of Barkhausen Noise in magnetic glasses based on a model Hamiltonian
that couples the mechanical properties of an amorphous solids to its magnetic degrees of freedom. In this Letter we announce a model that
contains long-ranged dipole-dipole interactions such that the magnetic domains appear labyrinthine (see Fig. \ref{laby}) in accordance
with the expectation that Barkhausen Noise will be associated with the movement of domain boundaries. Nevertheless we will report here
results that are quite surprising, in strong contradiction with many of the expectations in the field.
%%%%%%%%%%%%%%%%%%%%%%%%%%%%%%%%%%%%%%%%%%%%%%%%%%%%%%%%%%%%%%%
%%%%%%%%%%%%%%%%%%%%%%%%%%%%%%%%%%%%%%%%%%%%%%%%%%%%

{\bf The Model}: Our model Hamiltonian represents a binary glass with magnetic degrees of freedom \cite{12HIP,14HIPS,14GSP}:
\begin{equation}
\label{umech}
U(\{\mathbf r_i\},\{\mathbf S_i\}) = U_{\rm mech}(\{\mathbf r_i\}) + U_{\rm mag}(\{\mathbf r_i\},\{\mathbf S_i\})\ ,
\end{equation}
where $\{\mathbf r_i\}_{i=1}^N$ are the 2-dimensional positions of $N$ particles in an area $L_x\times L_y$ and $\mathbf S_i$ are spin variables. The mechanical part $U_{\rm mech}$ represents a standard binary mixture of 50\% particles A and 50\% particles B,
with Lennard-Jones potentials having a minimum at positions $\sigma_{AA}=1.17557$, $\sigma_{AB}=1.0$ and $\sigma_{BB}=0.618034$ \cite{09BSPK}. These parameters are known to provide good glass formation without crystallization. The energy parameters are selected as $\epsilon_{AA}=\epsilon_{BB}=0.5$ and
$\epsilon_{AB}=1.0$, in units for which the Boltzmann constant equals unity. All the potentials are truncated at distance 2.5$\sigma$ with two continuous derivatives. $N_A$ `A' particles carry spins $\mathbf S_i$; the $N_B$ `B' particles are not magnetic. Of course $N_A+N_B= N$. In the present model the spins $\mathbf S_i$ are classical Heisenberg spins in 3-dimensions; these can point anywhere on the unit sphere.

The magnetic contribution to the potential energy is chosen to allow the creation of labyrinthine magnetic domains
\begin{eqnarray}
&&U_{\rm mag}(\{\mathbf r_i\}, \{\mathbf S_i\}) = - \sum_{<ij>}J(r_{ij}) \B S_i\cdot\B S_j -   \B B \cdot\sum_i \B S_i \nonumber\\&&-  \sum_i K^{\parallel}_i\cos^2{(\phi_i-\psi_i(\{\mathbf r_i\}))}-K^{\perp} \sum_i S_{iz}^2 \nonumber\\
&&-\sum_{\langle ij\rangle} \frac{3(\B \mu_i\cdot\B r_{ij})(\B \mu_j\cdot\B r_{ij})-(\B \mu_i\cdot\B \mu_j)r_{ij}^2}{r_{ij}^5} \ .
\label{magU}
\end{eqnarray}
Here $r_{ij}\equiv |\mathbf r_i-\mathbf r_j|$ and the sums are only over the A particles that carry spins. The exchange parameter $J(\mathbf r_{ij})$ is a deterministic function of a changing inter-particle position (either due to affine motions induced
by an external strain or an external magnetic field or due to non-affine particle displacements, and see below).  We choose for concreteness the monotonically decreasing form $J(x) =J_0 f(x)$ where $f(x) \equiv \exp(-x^2/0.3)+H_0+H_2 x^2+H_4 x^4 $ with
$H_0=-5.51\times 10^{-8}\ ,H_2=1.68 \times 10^{-8}\ , H_4=-1.29 \times 10^{-9}$ \cite{13HPS,13DHPS}.
This choice cuts off $J(x)$ at $x=2.5$ with two smooth derivatives.   Finally, in our case $J_0=6$.
The second term is the interaction with the external magnetic field.
The next term represents the effect of an in-plane ($x-y$) local anisotropy, where
the local axis of anisotropy $\psi_i$ is determined by the local structure. The angle $\phi_i$ is determined by the projection of the
spin $\B S_i$ on the $x-y$ plane and is measured with respect to the $x$-axis. To find $\psi_i$ we define the matrix $\mathbf T_i$:
\begin{equation}
T_i^{\alpha\beta} \equiv \sum_j J( r_{ij})  r_{ij}^\alpha r_{ij}^\beta/\sum_j J( r_{ij}) \ ,\quad r_{ij}\equiv |\B r_i-\B r_j| \ ,
\end{equation}
where we sum over the particles that are within the range of $J( r_{ij})$. The matrix $\mathbf T_i$ has two eigenvalues in 2-dimensions that we denote as $\kappa_{i,1}$ and $\kappa_{i,2}$, $\kappa_{i,1}\ge \kappa_{i,2}$. The eigenvector that belongs to the larger eigenvalue $\kappa_{i,1}$ is denoted by $\hat {\mathbf n}$. The easy axis of anisotropy is given by $\psi_i\equiv \sin^{-1} (|\hat n_y|)$. Finally the coefficient $K_i$ is defined as
\begin{equation}
\label{KK}
K_i \equiv \tilde C[\sum_j J( r_{ij})]^2 (\kappa_{i,1}-\kappa_{i,2})^2\ ,~~ \tilde C= K_0/J_0\sigma^4_{AB} \ .
\end{equation}
The parameter $K_0$ determines the relative strength of this random local anisotropy term with respect to other terms in the Hamiltonian \cite{foot}. The next term in the Hamiltonian represents the perpendicular (out-of-plane) anisotropy in the $z$-direction. The last term is the dipole-dipole weak but long-ranged interaction;
here $\B \mu_i$ is defined as $\mu_B \B S_i$ where $\mu_B$ is taken as $\mu_B^2=0.1$.
We have chosen $B$ in the range [-0.65,0.65]. At the two extreme values all the spins are aligned along the direction of $\mathbf B$.

{\bf Barkhausen Noise}: The model has an obvious enormous parameter space with very many interesting effects that are beyond the scope of this Letter. Here we explore parameters that result in a labyrinthine pattern of ``up" and ``down" spins.
In Fig. \ref{laby} upper panel we show a snapshot of the magnetic domains of the present model when the external magnetic field is zero.
We reiterate that the parameters were chosen such that competition between the dipole-dipole interaction and the perpendicular anisotropy result in all the spins pointing either ``up" or ``down" in the $z$-direction. Having this structure with $B=0$ we next switch on a magnetic
field in the $z$ direction which we ramp up in small steps (quasi-statically), applying conjugate gradient energy minimization after
each such step to bring the system back to mechanical and magnetic equilibrium. The effect of the increasing magnetic field is
exemplified by the middle and lower panels of Fig. \ref{laby}; we observe the creation of new domains and the coarsening of the existing domains of ``up" spins at the expense of ``down" spins. The coarsening occurs by a movement of the domain boundary.

The creation of new domains and the movement of the domain boundary occurs in jerks (sometime referred to as ``avalanches") such that a number of spins $s$
flip from ``down" to ``up" when $B$ is increasing, and later, after saturation, when all the spins are pointing ``up" the opposite
changes occur when the magnetic field is decreased to the point of being negative. The flip of $s$ spins is equivalent to a change in magnetization $\Delta m\equiv  s/N_A$. A typical hysteresis loop exhibiting the sharp changes in the magnetization $M$ is shown in Fig. \ref{hyst}.
The magnetization curve has smooth sections punctuated by discontinuities whose size and distribution will be the focus of this Letter.

\begin{figure}
\centering
\includegraphics[trim=0cm 0cm 0cm 0.05cm, clip=true, scale = 0.35]{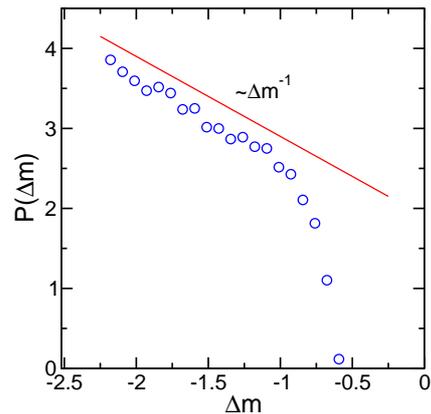}
\caption{A logarithmic binning of the measured data in our model with $N=2100$ for the probability $P(\Delta m)$ plotted in double-logarithmic scale. It is very easy to believe that the data is well represented by a power law multiplied by a cutoff function with $\alpha$ in Eq.~(\ref{expect}) being very close to $\alpha=1$.}
\label{wrong}
\end{figure}

{\bf Statistics of the Barkhausen Noise}: There exists a large body of literature that expects the statistics of Barkhausen Noise,
as well of many other serrated responses, to be modeled by a power law multiplied by a cutoff functions, i.e.
\begin{equation}
\label{expect}
P(\Delta m)={\Delta m}^{-\alpha} f(\Delta m) \ ,
\end{equation}
where $f(x)$ is falling off rapidly for large values of $x$. If one is convinced of the verity of this form, it is very easy
to mislead oneself to support it by the data. For example, in Fig. \ref{wrong} we present the measured data from a system
of 2100 particles for $P(\Delta m)$. The data was collected in logarithmic bins and plotted accordingly as $log_{10} P(\Delta m)$
as a function of $\log_{10} \Delta m$.
%%%%%%%%%%%%%%%%%%%%%%%%%%%%%%%%%%%%%%%%%%%%
The reader can convince herself that the plot appears to agree with the expected form Eq.~(\ref{expect}) with $\alpha\approx -1$. In the rest
of this Letter we will show that this is in fact incorrect, and that in the present example there is no power law whatsoever, notwithstanding
the apparent scaling presented in Fig.~\ref{wrong}.

\begin{figure}
\centering
\includegraphics[scale = 0.40]{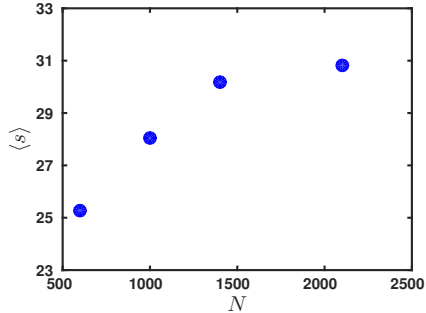}
\caption{The average $\langle s\rangle$ as a function of system size $N$. The indication is that $\langle s\rangle$
reaches a system-size independent limit as $N\to \infty$.}
\label{average}
\end{figure}

To understand what is the actual statistics in the present model we need to think what is happening when the magnetic field is ramped up or down. Indeed, the magnetization is changed due to flips of some number of spins from ``down" to ``up" when the
magnetic field is increased or from ``up" to ``down" in the opposite case. Our model here is not close to any apparent criticality,
so we should expect that there exist an average number of spins $\langle s\rangle$ that flip in an typical avalanche, and
that this average number does not increase like the number of particles $N$ when the latter is increased. In fact, we have measured
this average as a function of system size, cf. Fig.~\ref{average}, where it becomes clear that $\langle s\rangle$ tends to
a system-size independent value when $N\to \infty$. As far as one can see this appears to be the only constraint
of the statistics $P(s)$, together with the normalization condition
\begin{equation}
\sum\limits_{s=s_{min}}^{s_{max}} P(s)=1,
\label{norm}
\end{equation}
The average number of flipping spins is fixed by
\begin{equation}
\sum\limits_{s=s_{min}}^{s_{max}} s P(s)=\langle s\rangle.
\label{mean}
\end{equation}
%%%%%%%%%%%%%%%%%%%%%%%%%%%%%%%%%%%%%%%%%%%%%%
%%%%%%%%%%%%%%%%%%%%%%%%%%%%%%%%%%%%%%%%%%%%%%%%%%%%%%%%%

In accordance with the principle of maximum entropy \cite{J57} to find the actual distribution $P(s)$ we should maximize the information entropy
\begin{equation}
S=-\sum\limits_{s=s_{min}}^{s_{max}} P(s)\ln P(s),
\label{entr}
\end{equation}
subject to the constrains defined by
Eq.~(\ref{norm}) and Eq.~(\ref{mean}). The standard method of  Lagrange multipliers \cite{J57} is employed
with the final result
\begin{equation}
P(s)=\frac{e^{-(s-s_{min})/\langle s\rangle}}{\langle s\rangle}\ . \label{pofs}
\end{equation}
Note that if the reasoning leading to Eq.~(\ref{pofs}) is accepted, we have no free parameter in comparing
this prediction to our data, since in the present case $s_{min}=1$ and $\langle s\rangle$ is known for every system size
$N$.

\begin{figure}
\centering
\vskip .4 cm
\includegraphics[scale = 0.4]{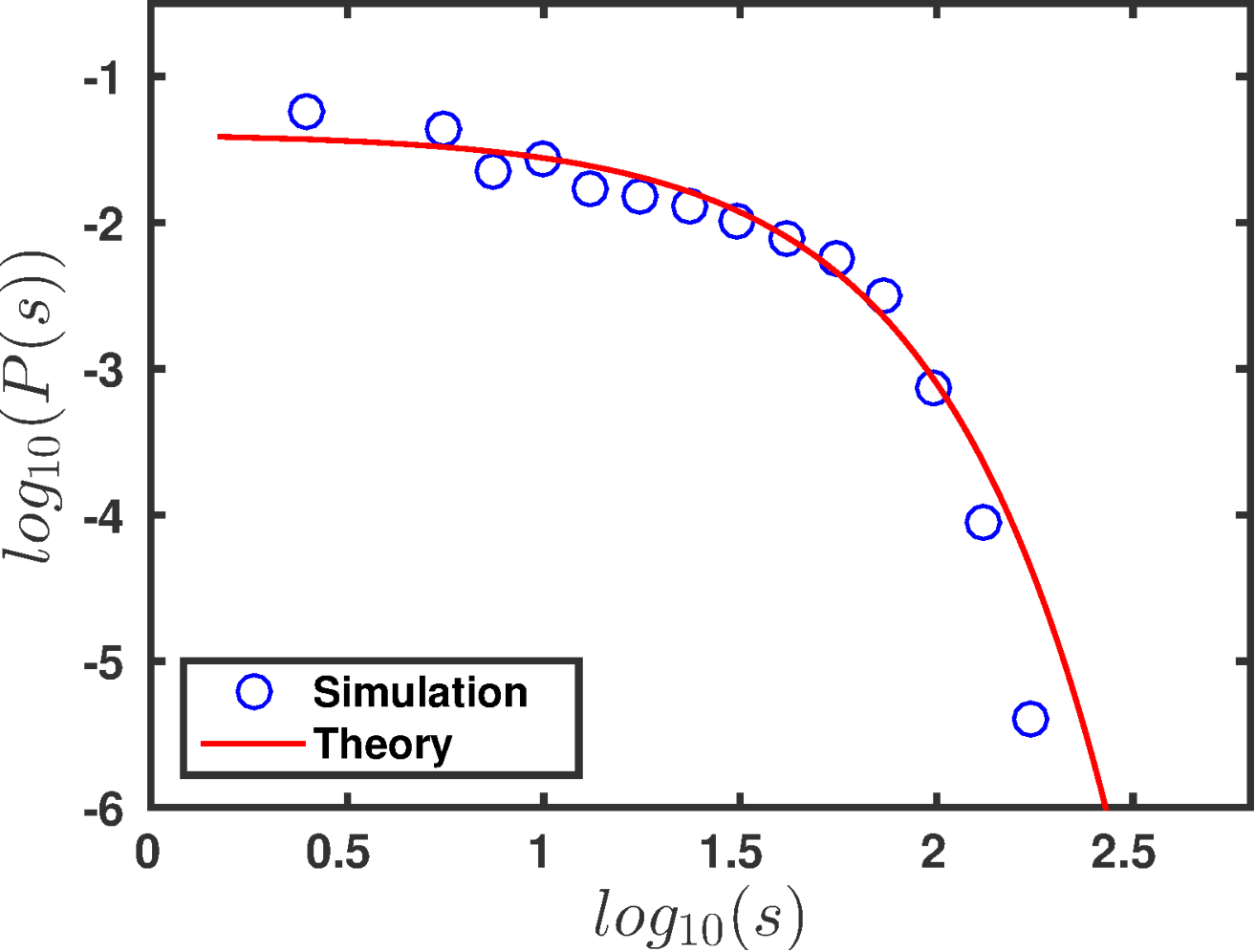}
\vskip .4cm
\includegraphics[scale = 0.4]{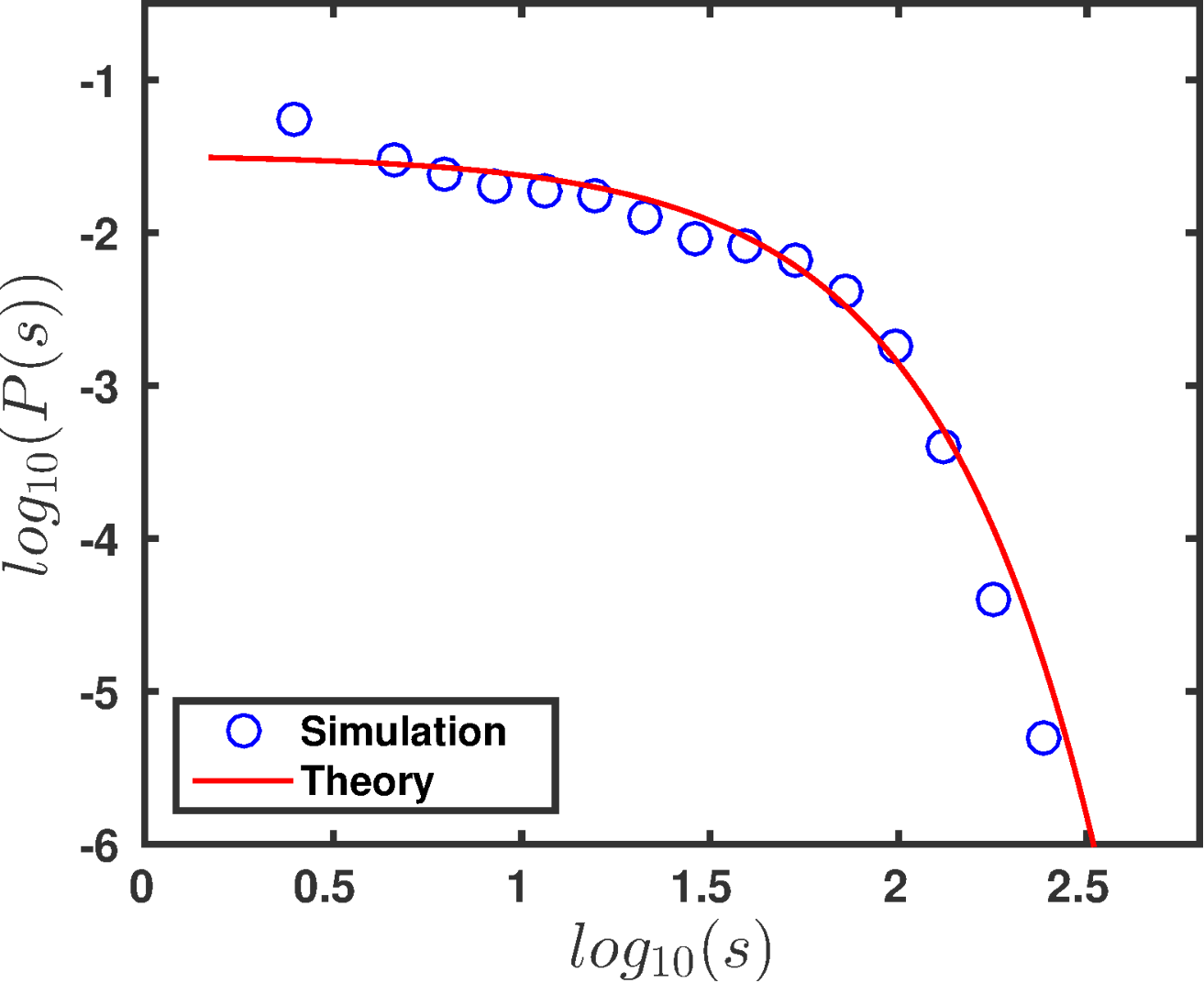}
\vskip .4cm
\includegraphics[scale = 0.4]{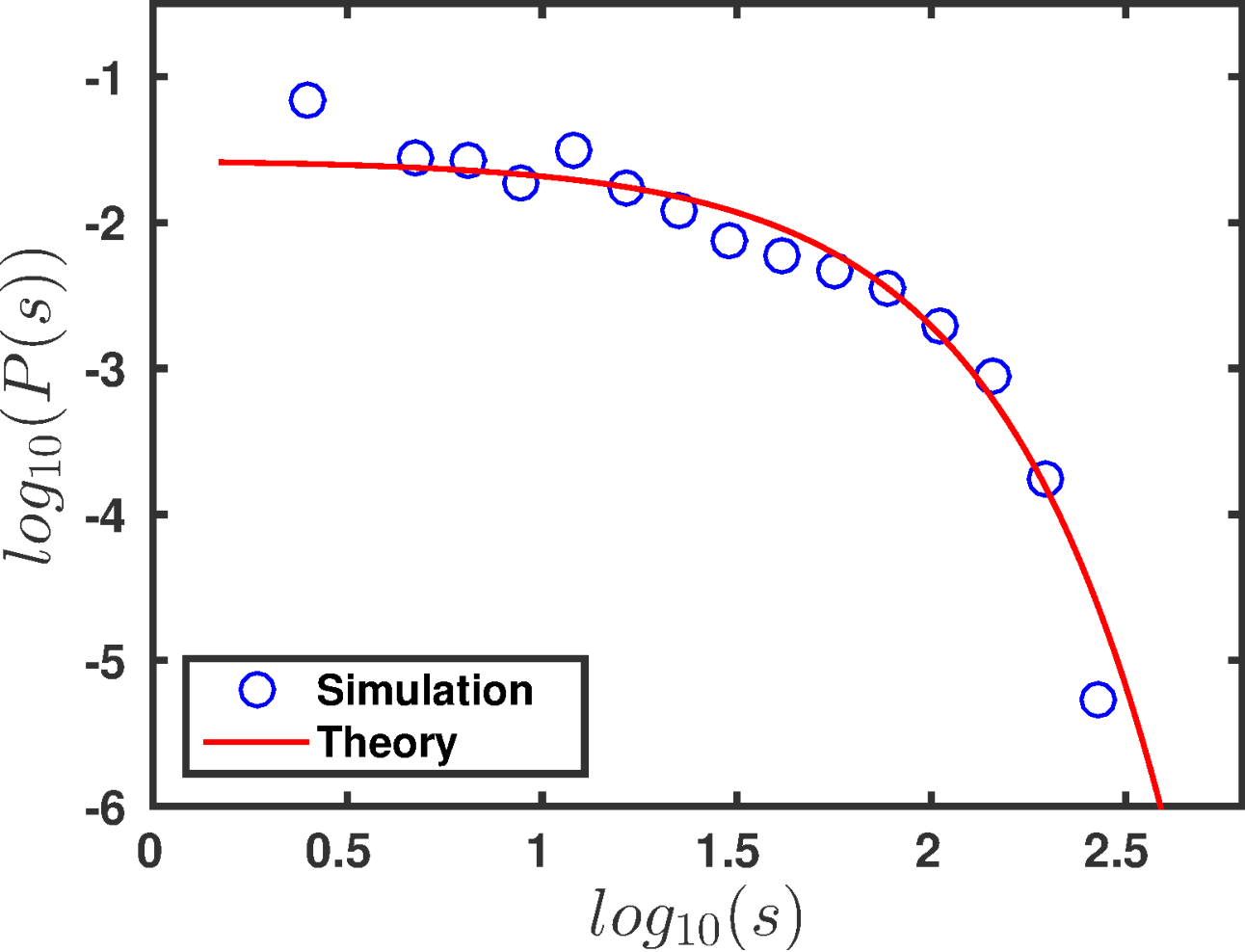}
\caption{Comparison of the statistics of the number of spin flips $P(s)$ to the theoretical formula Eq.~(\ref{pofs}).
Note that the comparison is done without any free parameter, showing an excellent agreement.
Upper panel: $N=600$; Middle panel: $N=1000$; Lower panel: $N=2100$.}
\label{compare}
\end{figure}

In Fig.~\ref{compare} we show the comparison of our data for three system sizes $N=600, 1000, 2100$ to the prediction Eq.~(\ref{pofs}) without any free parameter.
%%%%%%%%%%%%%%%%%%%%%%%%%%%%%%%%%%%
 Obviously the present theory appears to agree excellently well with the data, showing
that in the present model the Barkhausen Noise has very simple exponential statistics without any power law.

The fundamental reason why the dipole-dipole long range term does not bring us any closer to a mean-field
theory of the Barkhausen Noise is that the particles in this glassy system are allowed to move. This important change from crystalline models can be seen by simply flipping one spin in an equilibrated configuration and observing the effect of subsequent energy minimization keeping the flipped spin frozen.
In a crystalline model neighboring spins are bound to react, and the effect may or may not be long ranged.
In the present class of models the spin-carrying particles can rearrange themselves without changing their spins and reach a new energy minimum. The responding displacement field will decay slowly as a power-law,
but the magnetic response can be highly localized or non existent. This can be quantitatively tested in the
present model by computing the inverse Hessian matrix, $\B H^{-1}$. This operator is proportional to the
Green's or response function of the material \cite{14GLDLW}. In the present model the Hessian has entries for the
positional degrees of freedom $\B r_i$ and the spin degrees of freedom $\B S_i$. We have checked that as a function of $r_{ij}$ the positional entries of the inverse Hessian decay slowly, as a power law with an
asymptotic $1/r^2$ law. In contrast the spin entries decay very rapidly, (faster than $1/r^4$), explaining the non existence
of a divergence of $\langle s \rangle$ when $N\to \infty$.

{\bf Summary and Conclusions}: The present Letter indicates a number of important conclusions for the large community that is
interested in serrated responses. Barkhausen Noise and other similar phenomena can appear with statistics that vary enormously
depending on the underlying microscopic dynamics. Note that in the present case we considered the cleanest possible case with
temperature $T=0$ and without any mechanical strains or stresses, and yet the expected behavior Eq.~(\ref{expect}) did not
materialize itself. A second, and even more worrisome conclusion is that it is very easy to mislead oneself to present the
data without supporting theory in the form of Eq.~(\ref{expect}), and different methods of binning the data may lead to
different exponents $\alpha$ that are not really there. Finally, we conclude that the richness of behaviors that begins
to unfold itself with different microscopic models underlines the usefulness of such models - their simulation is straightforward,
the quality of the data is excellent and in general it is relatively easy to understand what is the nature of the serrated
response under study. We thus plan to continue along the lines presented here and study further universality classes of
serrated responses with the same care and precision.

{\bf Acknowledgments}:
This work had been supported in part by an ERC ``ideas" grant STANPAS. PKJ acknowledges a PBC fellowship from the Israel Council of Higher Education (VATAT).


\begin{thebibliography}{0}
\bibitem{foot}
The reader should note that the anisotropy term could be chosen in the form $\hat{\B n}\cdot \B S_i$ rather than being restriced
to the $x-y$ plane. This will result in a richer model that we plan to study in the future.
\end{thebibliography}
\end{document}